# Physical Biomodeling: a new field enabled by 3-D printing in biomodeling

Promita Chakraborty
QuezyLab, Sunnyvale, California 94085, USA.
promita@quezylab.com

## Summary

Accurate physical modeling with 3D-printing techniques could lead to new approaches to study structure and dynamics of biological systems complementing computational methods. Computational biology has become an important part of research over the last couple of decades. Now 3D printing technology opens the door for a new field, Physical Biomodeling, at the intersection of experimental data, computational biology and physical modeling for study of biological systems, such as protein folding at nano-scale. Here I explore this new domain of precision physical modeling and correlate it with existing visualization and computational systems and future possibilities. Dynamic physical models can be designed to-scale that can serve as research tools in future along with existing biocomputational tools and databases, adding a third angle to tackle unsolved scientific problems.

## Introduction

Biological structures, from the nano-scale to macro-scale are known to flex, fold, grow or shrink, and are constantly evolving from one form to another. The dynamics of these structures further add to the complexity. A remarkable example of complex dynamic structures is the protein folding problem. At present we have sophisticated computational models to study the folding process. But is that enough? In 2011 in an article, TIME Magazine reported the amazing success of online FoldIt gamers in solving enzyme structure that baffled scientists for decades [1, 2]. These gamers solved the problem, not as a scientific undertaking, but as a 3D puzzle. It highlights the findings that human beings are naturally good at spatial reasoning, methods in which computers struggle hard. The observation opens up new questions: Can we use dynamic, physical models as new research tools to study the dynamics of biological systems? Is it any better to have a precise scaled, self-folding models sitting at your desk? Can we use such models together with existing computational models and software tools? And more importantly, is it feasible to make such systems accurately?

Computer simulations provide a successful means to study complex biosystems, but they still lack a platform that will facilitate an intuitive understanding of the underlying complexity that is so crucial for deep insights leading to discovery. For example in protein folding, biocomputational models have limited themselves to modeling in terms of force fields, energy and entropy. The dynamics is thought of as entropy minimization problem. These are precise and successful methodologies for studying shapes and dynamics in the protein folding process. On the other hand, in physical models we represent dynamics using totally different set of parameters. Constraints for hard-sphere collisions, and the pathway for folding into a specific structure can be given as examples to show the approach in physical modeling. As a result, this different approach that handle the same problem from another angle, could bring forth a complementary set of data. The tangibility, geometry, degrees-of-rotation and rotational biases ingrained within physical models will make it easier to collect data on problems that study dynamics in biological systems.

In recent years, extensive studies have been made in CAD methodologies, solid modeling geometry and 3D printing techniques. But the biology research community is yet to gain from the advancements. A marriage of 3D CAD and biocomputation is in sight with cutting-edge technologies like 3D bioprinting, CAD-Nano and DNA scaffolds. These are transformative and disruptive technologies that society will greatly benefit from in future. Specifically, the advent of 3D-printing adds a new player to the game in physical modeling with the Design ⇒ Print ⇒ Play paradigm. The questions that come to mind is: What happens if we want to build macro-scale, precise physical models of any of these molecular systems? Can these models actually serve as instruments for future research – and if this is indeed possible, can we imbibe digital knowledge into them so that they can "compute" by virtue of the shape and dynamics? The Peppytide model (the example case study provided in this paper) is one of the initial efforts to establish the feasibility of such physical models for protein folding [3-6, 16].

Form-specific physical objects are easy to manipulate by hand, and can be designed to emulate the shape and mechanics of bio-systems. The interplay between the physical and computational models can be analyzed, evaluated and designed to contribute in a complementary manner, so that the two can work together to provide a more effective tool, e.g. mimicking the process of protein folding with your hand and parallelly generating its digital shadow. Thus there is a need felt for direct representation and manipulation of these dynamic shapes that can have a digital presence as well as physical, pulling from the strength of both.

A new paradigm emerges from this analysis wherefrom the form and flexibility of the real-world object can contribute to the better parameterization of the computer models, and to a deeper intuitive understanding of the biological phenomena. Thus this paradigm sets forth new ways to think about bio-systems that complement the existing methods.

## A New Field

*Physical Biomodeling* is a new area of exploration at the interface of computer science, precision physical models and biological systems. While tremendous advances have been made in computational biology, the cutting-edge 3D printing and other makerspace tools and materials provide unprecedented opportunities for a third angle into the landscape, thus uncovering a new computational space for modeling and prediction that has remained unexplored so far (Figure 1). This approach is especially important as it provides a new way to look at the old scientific problems, thus creating opportunities to produce new, unexplored sets of data.

To study the relationships between different types models and data, I define 3 sets as:

- N: experimental data from natural systems in native states
- C: existing computational models
- P: form-specific, dynamic physical models

The relationship between these sets tie together the concepts of form-specific physical-digital interfaces and precision physical models that are dynamic and true to their functions. There are 6 processes connecting these 3 sets (Figure 1). The sets and processes together form the basis of the philosophy behind the field of Physical Biomodeling. Through these relationships and mappings, a new computational paradigm emerges for physical-digital interfaces for studying of biological phenomena (e.g. protein folding timesteps) that focus on shape and dynamics. New opportunities for scientific exploration emerges at the intersection of these relationships, providing us with unexplored areas of study that might complement the existing bio-computing frameworks. These possibilities and scopes are discussed in detail at a later section.

Through exploring Processes 1-6, we see that a relationship-triangle exists between the experimental data from natural systems and native states (N), the computational models for biosystems (C), and scaled, accurate physical models of biosystems (P). Processes 1 and 2 have already existed in the literature for a long time. Physical Biomodeling brings forth Processes 3-6 that could provide a new way to look at the old problems in biology. We arrive at a computational space at the intersection of N, C and P that has so far remained unexplored because of the difficulty in designing and fabricating accurate, scaled physical models of biosystems. Further, as discussed later, no such bio-cum-CAD platform exists yet to facilitate this union. Hence, we have opportunity to develop a new class of software tools for biology to bridge this gap.

The concepts laid down for Physical Biomodeling is generic and can be applicable to any scientific phenomena. As an example case study, the six processes and their utility have been illustrated for the Peppytide model in the following sections for the problem of protein folding.

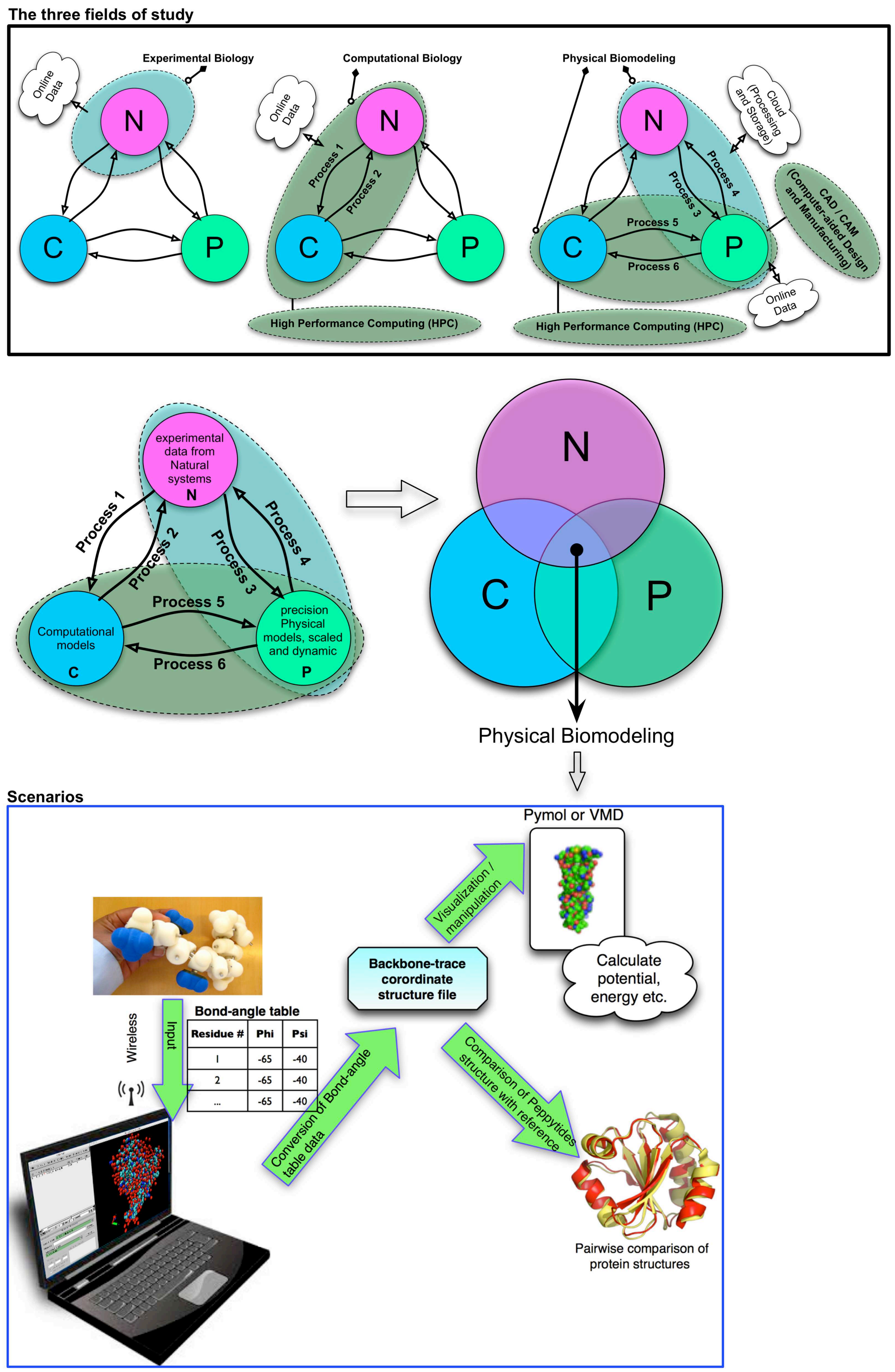

**Fig. 1. A new field.** Physical Biomodeling provides a new angle to study existing problems, a new field of study at the intersection of N, C and P: the Entity-Relationship diagram in the physical-digital computing space. N: experimental data from Natural systems, C: Computational models, P: precision Physical models.

## The Physical Model Parameter Space

Biological systems involve complex phenomena, and encapsulating these characteristics within a physical body is thought to be difficult. For example, representing the polypeptide chain, a generalized protein chain, along with its complex degrees-of-freedom, by a physical scaled model that will fold dimensionally-accurately retaining protein folding behavior, was thought to be quite difficult before the Peppytide model proved otherwise [3,4,17]. Now with 3D-printing technologies and other maker tools, and possibilities for CAD-cum-biocomputation platforms, we are poised to explore this new domain of study.

Physical models have a completely different set of parameters than computational models. This is another reason to study precision physical models as scientific tools. The parameters for Peppytide, the physical protein model, are given as an example: (1) hard-sphere size of atoms of elements (=0.7 * Van der Waal's radius), (2) the scaling factor (1Å = 0.3676"), (3) the magnet positions for rotational biases for bond rotations, (4) the magnet strengths calibrated to molecular forces, (5) weight and effects of gravity on folding process. My prior work has established that with only these parameters one can arrive at the precisely folded protein structures of amino acid chains up to 50AA[1] long. For example, the alpha-helix of 13 amino acids measured 6.75" (equivalent to 18.362 Å), while an alpha helix of same length from X-ray crystallography data of GCN4 Leu-zipper measured 18.4 Å [3,4]. Thus, two different sets of parameters through two different types of models, computational versus physical, arrive at the same structural state through different means. It will be interesting to now study how these two parameters map to each other. This concept forms the basic principle of Physical Biomodeling.

The mapping functions and their understanding will enable conversion of data between these the computational and physical models back and forth. For example, someone would be able to select a sequence of amino acids and compute the corresponding physical model parameters, and then 3D-print a precisely foldable particular sequence in order to study it. I discuss this mapping in more detail in Processes 5 and 6 below.

## Peppytide: An example case study for the Physical Biomodeling paradigm

Peppytide presents itself as an example proof-of-concept of the Physical Biomodeling principle and conceptualization. It is a 3D-printed physical model of polypeptides with flexible backbone that can be folded into various secondary and tertiary structures (Supplementary Figure S1). It was designed based on data from X-ray crystallography structures as well as established radii, bond distances and bond rotations. These models are scaled and comparable to experimental data and structures (Supplementary Figure S2).

**Processes 1 and 2.** These are the two traditional process involved in simulation, modeling and prediction. Process 1 involves fitting experimental data to mathematical and computational models, while Process 2 entails using the model to predict system behavior. For example, numerous studies have been done in attaining precise and fast computational models and folding principles using template matching, force field calculations and other methods [7-11] (Process 1), as it is an extremely important topic that has implications for medicine and drug design in addition to understanding the fundamental rules of protein folding. Extensive studies have also been made within the last two decades in designing small de novo proteins with less that 40% homology to known sequences to achieve a preferred structure [12,13] (Process 2). This body of work has firmly established the relationship between N and C (Figure 2) with the protein folding data available from X-ray crystallography and NMR structures of native states, and the *de novo* protein folds.

[1] AA: Amino Acid

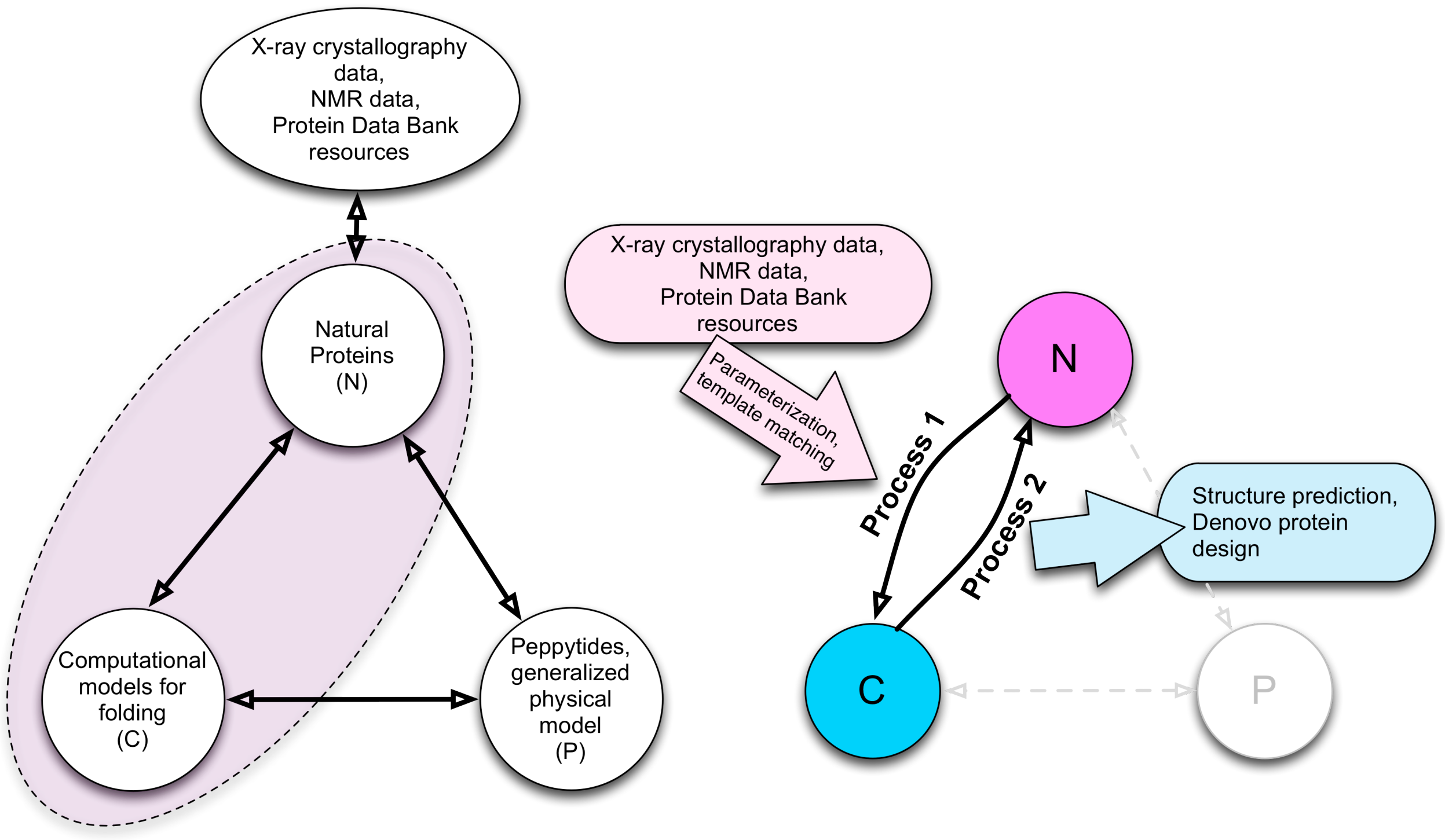

**Fig. 2. Processes 1 & 2.** A look at computational modeling based on existing observations and current concepts.

**Process 3.** This process, from N to P, explores the design cycle for focusing on the meticulous design of exact-scale physical models that address open problems in biology, bringing with it a new way to look at the same problems. The spatial, tangible, geometric approach to looking at a problem that has been extensively explored computationally and experimentally, would bring in fresh aspects – the third angle – to our knowledge and would feed-forward into the computational space. Even though there has been extensive developments in 3D-printers and CAD modeling techniques within the last 10 years, the idea of designing physical models for biosystems that demand precision, exactness and dynamics, is still daunting. Moreover, prototyping such models are themselves hard problems. But recently the feasibility of this concept has been established. For example, Peppytide which is one of the first such models that focus on the protein folding problem, demonstrates the exact-scaling, accuracy in degrees-of-freedom and the ability of the model to form various folded secondary and tertiary structure motifs found in proteins at their native states [3,4] (Figure 3, 4). This is a significant development that proves the feasibility of the approach of physical modeling and computation for scientific problems in biology that focus on structure and dynamics.

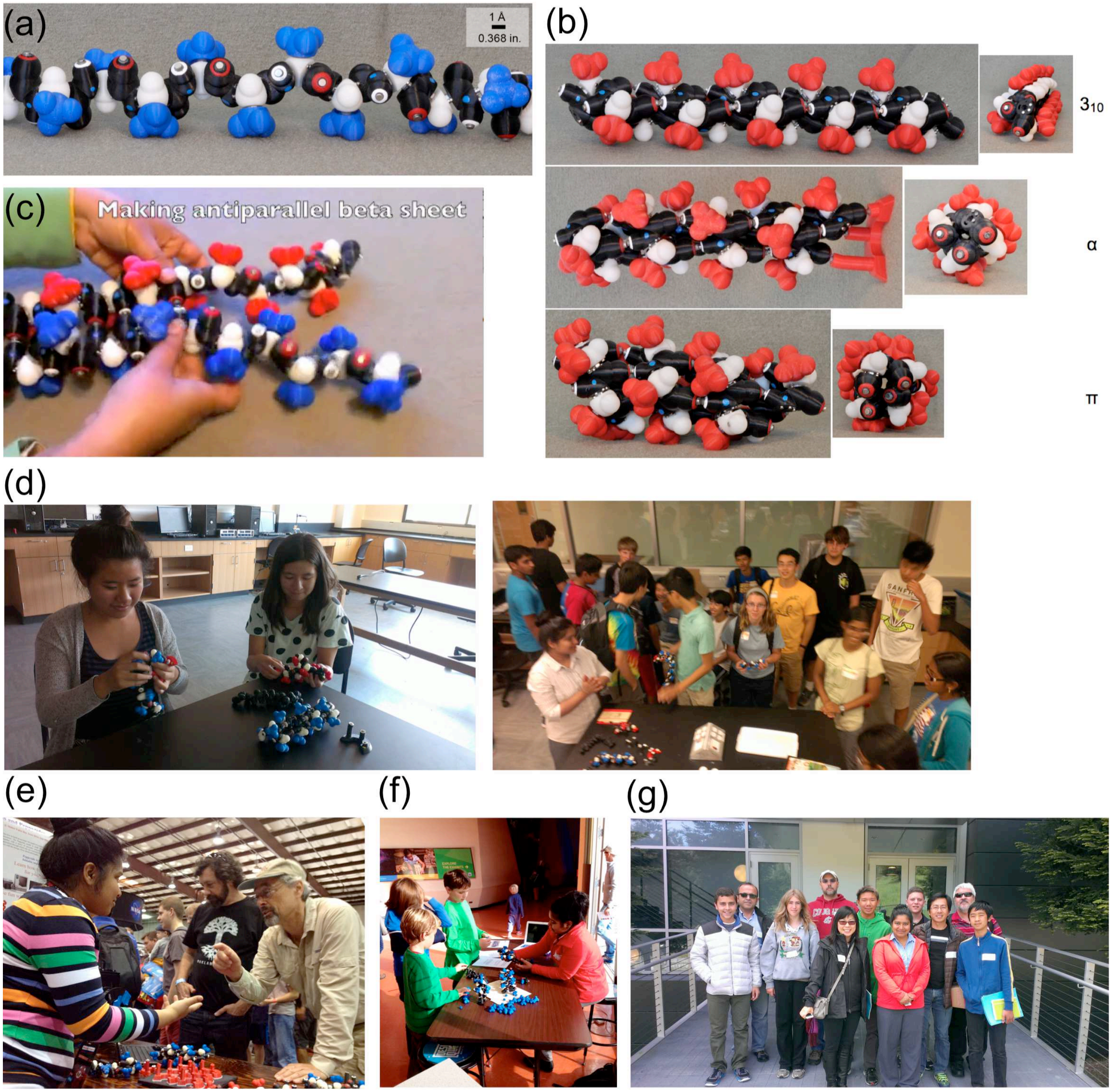

**Fig. 3. The Peppytide model and protein folding.** (a) A scaled, dynamic physical model of the polypeptide chain folded into various secondary structures found in proteins, (b) a comparison of the model folded into 310 helix, looser alpha helix and loosest pi helix, demonstrating the realistic flexibility of model backbone, (c) making antiparallel beta sheet, (d) demonstrating folding of secondary structures in Summer NanoCamp at Foothill College, 2014, (e) demonstrating protein folding with Peppytide at Maker Faires (2014, 2015, Bay Area, and National Maker Faire 2015, Washington DC), (f) demonstrating folding at Lawrence Hall of Science, UC Berkeley, 2013, (g) protein folding with participants from Johns Hopkins Center for Talented Youth at Lawrence Berkeley National Laboratory, 2014 (photos with permissions).

**Process 4.** Once a physical model is prototyped, it is crucial to test the performance of the model through rigorous testing and by comparison with established data. This validation establishes the relationship from P to N forming the process 4 testing and design cycle. As a use case of the process, the polypeptide chain model Peppytide was tested by folding it manually into a variety of secondary structures, like the most frequently found α-helix, β-sheets and β-turns, and the less frequent $3_{10}$ helix and π-helix (Figure 3). A few tertiary structure motifs (ββα, Osteocalcin ααα) were also made (Supplementary Figure S1). The models were then measured by hand with a ruler, and converted to nano-meter with the precise scaling factor (1Å=0.368"). Comparison with corresponding X-ray crystallography native state structures show good agreement, and hence the extreme precision of the model (Supplementary Figure S2).

The successful testing implies that the structures that can be made by hand with the model are possible structure-candidates for native-state protein conformations and vice versa (Figure 4). Thus there is an equivalence that exists between experimental data and physical models with their own set of parameters. Though these parameters (radius, size, scale, weight) are different from the parameters in the computational models, an analogous scenario can be envisioned that goes back and forth between N and P (analogous to N and C for computational biology). The scaled model lets anyone to get an accurate idea of dimensions of these structures at nano-scale, just by measuring the model with a ruler and using the conversion-factor, which would serve as the basis for the model being used for prediction in future. A comparison of Ramachandran plots of Peppytide with native state shows how close the model relates to protein behavior and what might be improved (Supplementary Figure S3).

Processes 3 and 4 demonstrate that a bijective mapping can exist between N and P for biosystems. It is a very crucial relationship in physical modeling of biological systems. For successfully modeling future physical biosystems, it is the necessary and sufficient condition to satisfy this relationship.

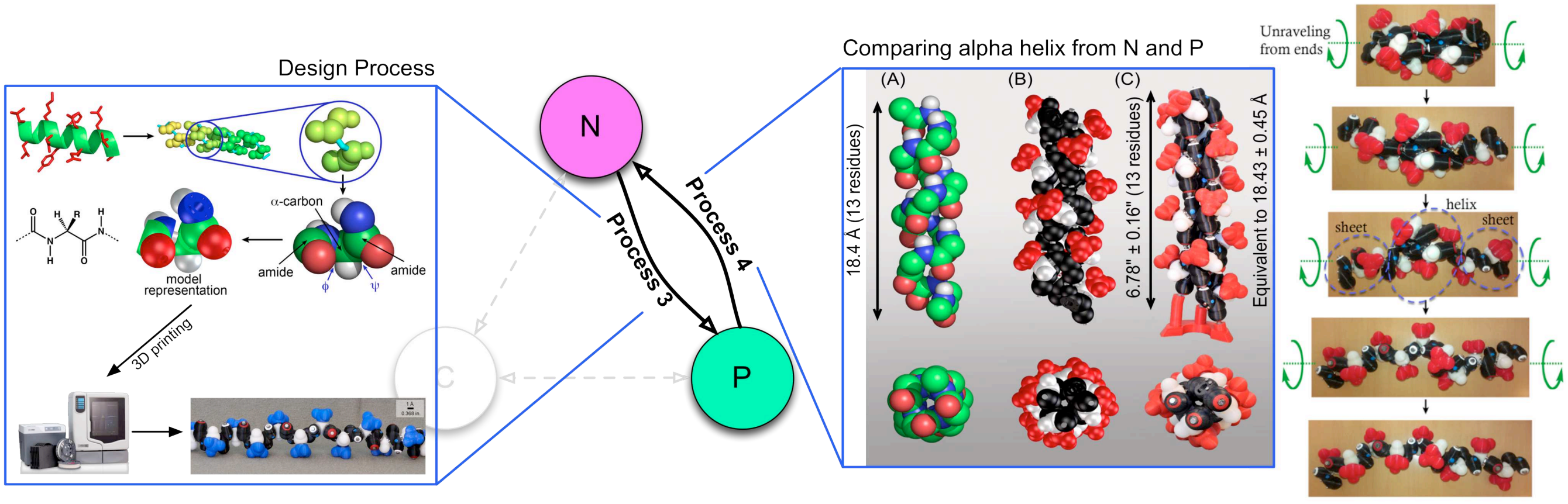

**Fig. 4. Processes 3 & 4.** Designing and validating physical models of polypeptide chain.

**Processes 5 & 6.** Tying together precision physical models with computational models is another hard problem. Firstly, no bio-CAD modeling platform currently exists that can easily map the digital representation of such a complicated physical model with the existing biocomputational platforms. Conversely, tracking a dynamic physical model with large degrees-of-freedom through cameras (or otherwise) real-time is a challenge. There is no existing computational framework that can make a smooth transition between computational and physical models back and forth. Some questions that can be raised here are: Can we strive for a paradigm like this: *structure determination* ⇒ *human-scale abstraction* ⇒ *3D- print* ⇒ *fold*? How would we solve the problem of automatically converting the standard .PDB files into physical foldable models of proteins? Do we need to decide on standards?

Thus, we have immense potential and lots of scope here to develop CAD-cum-biocomputational software tools and models to initiate the dialog between these two types of modeling. This is a yet-unexplored new class of software platforms and frameworks with lots of potential uses. Marrying CAD-based platforms and 3D-printing with biocomputational simulations opens up new possibilities for research in biology, and would eventually establish new grounds with Physical Biomodeling techniques. New standards for file conversions between CAD and computational representations also need to be established as a first step towards designing such platforms.

A new exploration is needed to establish the ground rules for designing such a digital-physical interfacing platform for polypeptides. In the preliminary plan of building such computational systems, I aimed to convert the elements of the physical model design into a digital-representation, for 3D-printing as a first step (Figure 5), using the Cyborg Project, the forthcoming CAD-enabled generic biocomputation framework, beta-released by Autodesk Inc. [14]. Cyborg has software features and modeling languages platform to program the digital-representation for polypeptides that would serve as the underlying knowledge base, but still no system exists to tie these different directions together. Thus these frameworks need to be built from scratch. Eventually, we hope to achieve an easy conversion of the design principles between CAD-oriented modeling and that for biological systems.

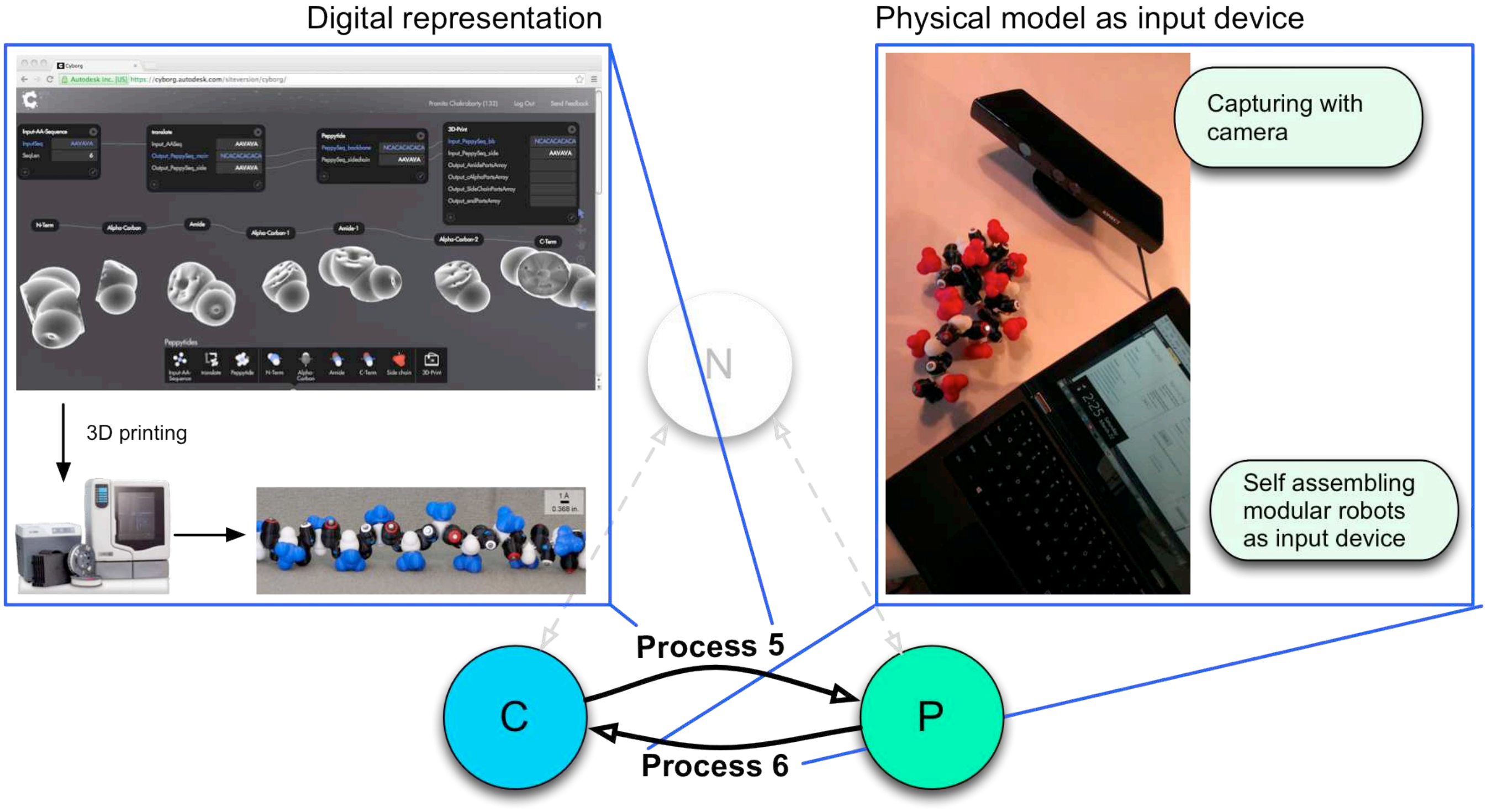

**Fig. 5. Processes 5 & 6.** The Digital-physical Interfacing.

A natural next-step would be to capture the structure of the physical models which can then be shadowed digitally in real-time with motion (Process 6). Similar kinds of object tracking experiments have been made recently with simpler objects like lego blocks using a Microsoft Kinect camera with depth-sensing feature, where the camera tracks the assembly, break-down and change-of-form of an object from its constituent lego parts on-the-fly [15,16]. It is still a long way before we can achieve such operations with biomodels that have so many degrees-of-freedom.

**The next step in protein models:**

Peppytide is a poly-alanine model. However, to do any kind of protein function-related study with physical models, we need it to be able to make any given sequence of amino acids as needed. Hence in the next step of development I added interchangeable side chains (Figure 6) leading to Peppytide-2 and PeppyChain. Peppytide and PeppyChain models have been submitted to NIH 3D Print Exchange database [17,18].

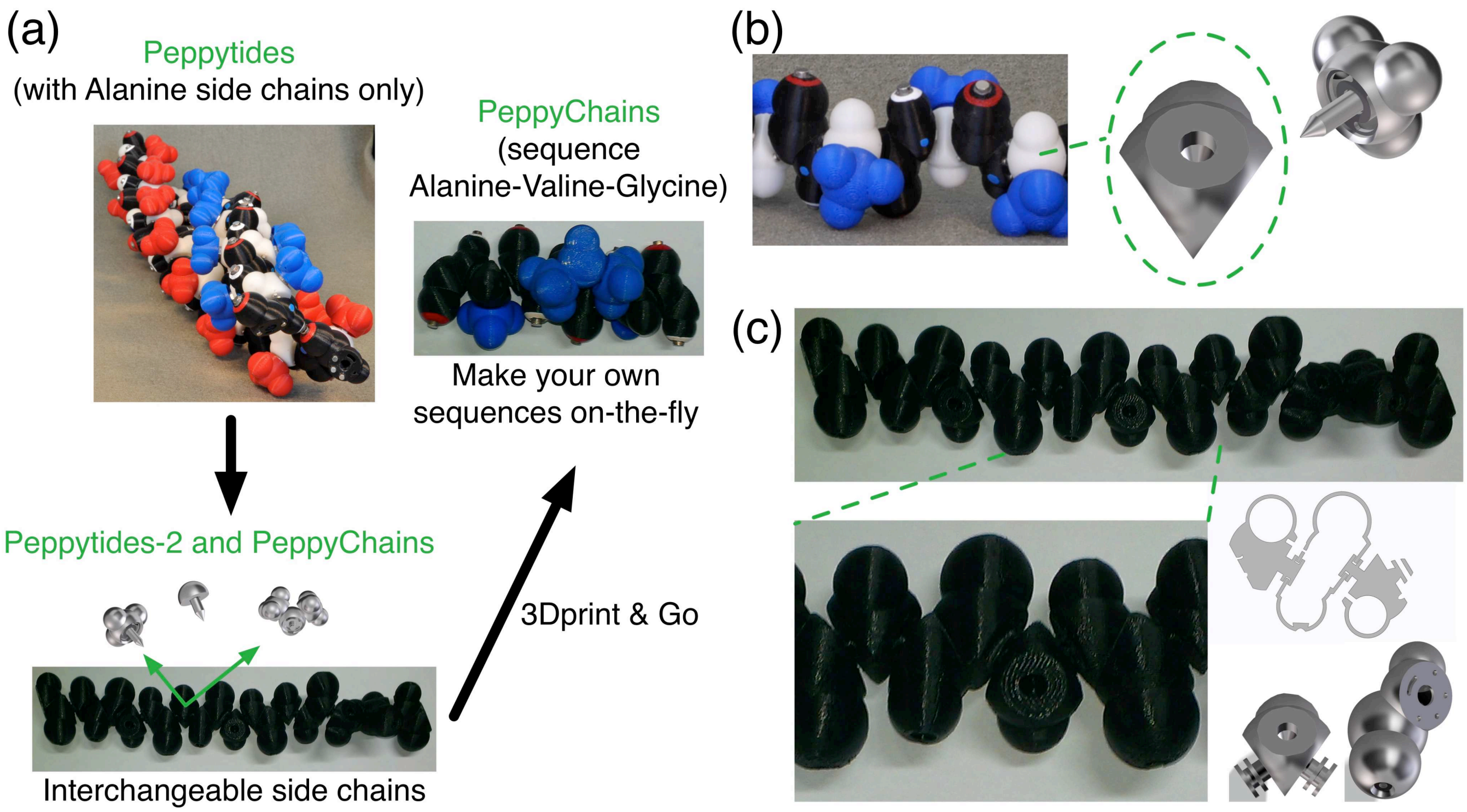

**Fig. 6. Interchangeable plug-in-out side chains for amino acid residues.** (a) Making variable amino acid sequences, (b) Peppytide 2: alpha carbon unit have been redesigned to enable plug-in-out side chain units, (c) PeppyChain 3D printed as a single unit preserving the degrees-of-freedom.

**Possibilities for scientific exploration in this computational space**

The physical modeling technique along with advances in 3D-printing, opens up new possibilities and challenges. Here I outline a few possibilities that can be built upon the foundation that has been laid down in this paper.

**(a) A viable input device for biocomputing.** Manipulating models with hand making desired shapes and then conveying that information to the computer, can give us a powerful tool for initial setup of conformational details for algorithmic analysis. It can also be used to study step-by-step change in conformations.

**(b) A viable output device for biocomputing.** Incorporating rotational degrees-of-freedom in physical models is still cumbersome with even the lightest of servos, but alternative technologies like shape-memory alloys and pneumatic devices might help. One interesting application of having such an output device in Peppytide project might be to make the model “display” the folding pathway as a function of time, given the ability to self-fold through actuators. Another use-case could be to have two remotely located models interacting with each other, with one model’s configuration transmitted to another, which could then fold accordingly.

**(c) Designing platforms for other or bigger biosystems.** The same principles can be extended to model the entire Central Dogma of Biology (DNA to RNA to proteins). Prior to Peppytide, I worked on the initial mock prototypes for the dogma (including transcription, translation, wobble[2] hypothesis) in BioTable (Figure 7), which included a computer screen as an interactive table surface with a top-mounted camera for object recognition, and paper, wood and bead models of nucleic acids, DNA, tRNA, amino acids and primary sequences of proteins [19,20,21]. BioTable was the precursor to and the big picture for designing the foldable, scaled Peppytide later on. Besides the protein folding problem, there are a lot more to focus here. There is scope for design of a lot of physical models within this information-space. Some examples on dynamics, folding and molecule-molecule interaction that would benefit from Physical Biomodeling principles are: DNA-mRNA interaction at the atomic level, tRNA folding, the codon-anticodon-tRNA mechanism and wobble hypothesis.

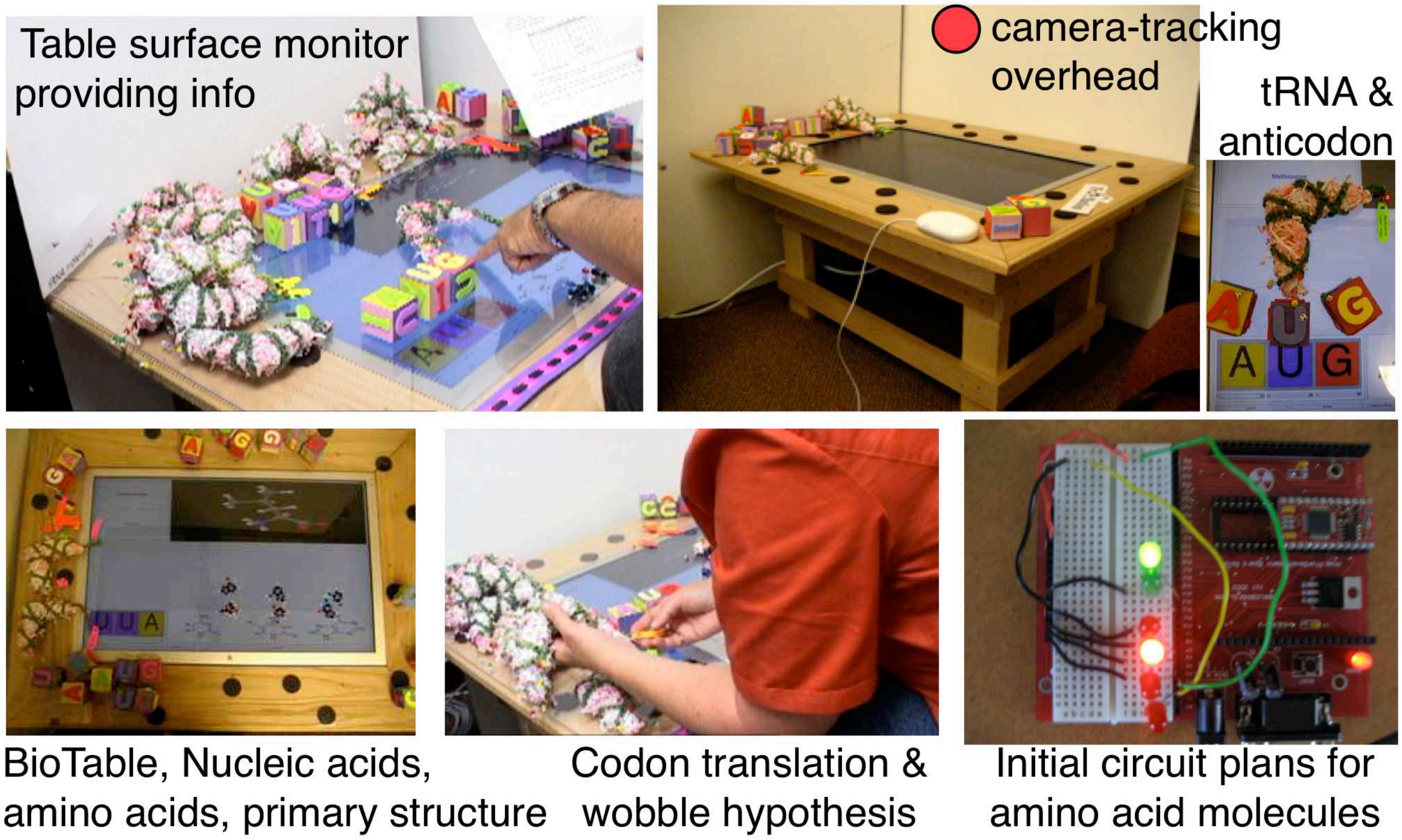

**Fig 7: BioTable.** Preliminary sketches, implementations and ideas for an interactive session with physical manipulatives (DNA to proteins), 2009-2011.

[2] A hypothesis explaining why or how multiple codons can code for a single amino acid.

**(d) Online NIH 3D-Print Exchange Database as a viable database for future Physical Biomodeling researchers.** Over the years, the Online Protein Database (http://www.rcsb.org) has become a crucial repository for researchers working with protein sequences and structures. The database serves as a concise citable site and a dedicated storage space for 3D structures from experiments and computer models, each having a unique PID#. Likewise, research in Physical Biomodeling would generate model that could be publicly made available through NIH's effort for 3D-printable hand-held model repository, called the NIH 3D Print Exchange (http://3dprint.nih.gov/). Currently, the website hosts various 3D-printable model files (STL and X3D files) for biomolecules or other biological systems that are scientifically accurate or medically applicable. Each model has a unique identifier (Model ID#). The Model ID for Peppytide and PeppyChain are 3DPX-001596 and 3DPX-001084 respectively [17,18] (figure 8).

In future, the database might also extend to host models that are computationally-aided, and have meta-data of the dynamics. In that case the models would not only be 3D-printable files or convertible to such files, but have to have capability to hold model meta-data and physical parameters as well. As an example, magnet positions for phi and psi can be parameterized to generate a 3D-printable specific pre-calculated structure for a particular protein or peptoid. Like the .PDB, .PSF or .STL file standards, new standards for files need to be developed at the digital-physical-biological interface that would enable such parameterization of dynamic physical models, and is essential to support usage of a common database across the entire research community for Physical Biomodeling. For example, for proteins, the standards from the .PDB files and the CAD-model files can be united into a single file format to contain the necessary information from both worlds, and for easy conversion between the physical and the digital.

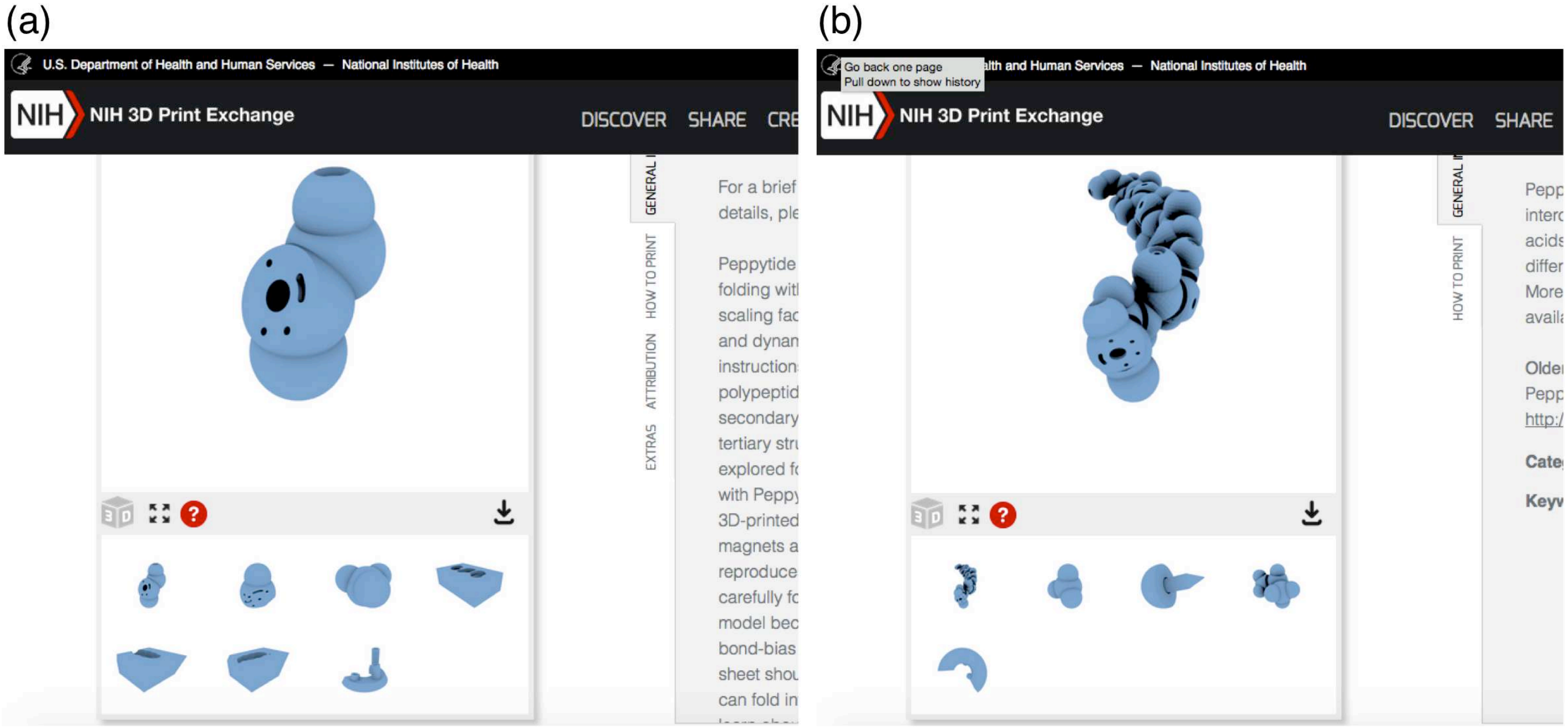

**Fig 8: Models at NIH 3D Print Exchange Database.** STL files and instructions to assemble / make models. (a) Peppytide [Model ID 3DPX-001596], (b) PeppyChain [Model ID3DPX-001084], [17,18].

## New frontiers for protein folding with physical models:

A few exciting and important possibilities, particularly for research in the protein folding problem is as follows.

**(a) Exploring de novo protein structures.** Designing de novo structures of proteins is a time-consuming process with computational predictions followed by experimental verifications. A scaled model can act as an in-between to speed up the cycle and might help in quicker rejections of false positives.

**(b) Study of misfolded proteins and aggregates.** Proteins often misfold and form aggregates, but have specific structural traits. An important problem would be to explore the behavior of insoluble proteins using physical models, to pinpoint the idiosyncrasies of misfolded ones like β−amyloids. By easily changing side chains, the physical models could provide a quick, initial study on the effects of mutation in protein folding without extensive calculations. For such studies to be most effective, we need to provide a means for the physical model to send its conformational information to the computer for further processing.

**(c) Electrostatic and hydrophobic interactions.** These forces are critical in protein folding. Implementing electrostatic effects and hydrophobicity in a scaled, realistic way with respect to other interactions, is the key to making stable tertiary structures using a physical model. The concepts in simulation platform sketched as example Process 5 above, might be eventually extended to develop simulations of physical models to computationally plan out a hydrophobic design scheme for the physical models. These interactions need to be orthogonal to the hydrogen bonding forces.

**(d) Self-folding of protein models.** Currently there is a growing interest in developing self-folding modular robots that can perform a particular task. The intersection of microelectronics, pervasive computing, and growing interests in biolocomotion have paved the path for self-folding 2D and 3D objects. Posey [22, 23], Moteins [24], Programmable matter [25] are a few examples from this class of objects. It is time to borrow these concepts to develop self-folding biopolymers.

**(e) Exploring other types of polymers.** The concepts of models of polypeptides can be extended to other polymer chains like poly-nucleotides, β−peptides, biomimetic peptoids, and industrially important polymers like Kevlar and polystyrene. Other options for synthetic polymers are conducting polymers and polyethyleneoxide. With exciting developments in material science and new materials, the possibility is endless here.

## In a nutshell

Over the last 50 years, bioinformatics and computational biology has revolutionized the way we do research in biology. Since the mid-twentieth century, increasing amount of data from experiments, biological sequencing, structure determination and other studies eventually fueled and outlined the developments in these two disciplines. Addressing these problems have in turn fed forward to developments in various branches within computer science including algorithms, graph theory, computational modeling, parallel computation methods, pattern recognition and visualization. The prevalence of visualizing tools repeatedly highlights the eternal human need for spatial understanding, to orient their insights in terms of geometry. However, the current biocomputational approaches leave tangibility and geometry at the periphery. With the latest developments in 3D-printing and CAD technologies it will now be possible to explore the convergence of these domains with biocomputation.

Moreover it is possible to extract the guiding rules, as learned from the Peppytide series of models, in order to use them in other problems in biology that focus on shapes and dynamics. These investigations can eventually define the boundaries of a new field of exploration at the physical-digital interface for biology, where the physical models and computational ones complement each other to enable better insights into the fundamental principles of biological systems.

The idea of embodying the physical and, most importantly, mechanical information of molecules directly into the artifact itself and to let it move at its own accord, was the underlying motivation behind the work with Peppytides, and is exportable to other complex systems. This point-of-view of looking at the problem is a substantial shift from having a representative wood block or other manipulatives as a tactile handle to direct the digital environment or augmented reality. Next, a way to successfully gather these physical quantities, like position and orientation of different subparts, directly from the artifact would pave the way for innovative ways of biomodeling in future. Such futuristic tools have the potential to change user behavior within the structural biology community, the same way as mouse and new visualizing tools disrupted and changed design of biocomputation tools a few decades ago.

**References**

[1] Matt Peckham. Foldit Gamers Solve AIDS Puzzle That Baffled Scientists for a Decade. Time Magazine, Gaming & Culture, 2011.

[2] Khatib, Baker et al. Crystal structure of a monomeric retroviral protease solved by protein folding game players. Nature Structural & Mol Biol, 18, 2011.

[3] Promita Chakraborty. A Computational Framework for Interacting with Physical Molecular Models of the Polypeptide Chain. PhD Dissertation, 2014.

[4] Promita Chakraborty and Ronald N. Zuckermann. Coarse-grained, foldable, physical model of the polypeptide chain. Proc Natl Acad Sci USA, 110(33):13368 –13373, 2013.

[5] Promita Chakraborty and Ronald N. Zuckermann. MAKE Projects: Peppytides. http://makezine.com/projects/peppytides/. [Online; accessed 30-Apr-2015].

[6] Peppytide. http:// www.peppytide.org [Online; accessed 30-Apr-2015].

[7] Ken A. Dill and Justin L. MacCallum. The protein-folding problem, 50 years on. Science, 338(6110):1042–1046, 2012.

[8] Ken A. Dill. Dominant forces in protein folding. Biochemistry, 29(31):7133 –7155, 1990.

[9] Christophe Schmitz, Robert Vernon, Gottfried Otting, David Baker, and Thomas Huber. Protein structure determination from pseudocontact shifts using ROSETTA. J Mol Biol., 416(5):668–677, 2012.

[10] Promita Chakraborty, Shantenu Jha, and Daniel S. Katz. Novel Submission Modes for Tightly-Coupled Jobs Across Distributed Resources for Reduced Time to-Solution. Phil. Trans R Soc A, 367(1897):2545–2556, 2009.

[11] Promita Chakraborty, Shantenu Jha, and Daniel S. Katz. Distributed Molecular Dynamics Simulations: A Case Study of Co-scheduling on the TeraGrid and LONI. TeraGrid 2009, Arlington, Virginia, USA, Jun 22-25, 2009.

[12] Bassil I. Dahiyat and Stephen L. Mayo. De novo protein design: Fully automated sequence selection. Science, 278(5335):82–87, 1997.

[13] Birte Hocker. Structural biology: A toolbox for protein design. Nature, 491(7423):204–205, 2012.

[14] Project Cyborg. http://www.autodeskresearch.com/projects/cyborg.

[15] Andrew Miller, Brandyn White, Emiko Charbonneau, Zach Kanzler, and Joseph J. LaViola Jr. Interactive 3D model acquisition and tracking of building block structures. IEEE Transactions On Visualization And Computer Graphics, 18(4):651–659, 2012.

[16] Ankit Gupta, Dieter Fox, Brian Curless, and Michael Cohen. DuploTrack: A realtime system for authoring and guiding Duplo block assembly. In Proceedings of 25th ACM Symposium on User Interface Software and Technology (UIST), 2012.

[17] Promita Chakraborty. Peppytides: 3D printed, scaled, foldable model of the polypeptide chain for protein folding. Model ID 3DPX-001596. NIH 3D Print Exchange. http://3dprint.nih.gov/discover/3dpx-001596 [Online; accessed 6-Jul-2015].

[18] Promita Chakraborty. PeppyChain: Peppytide protein models with interchangeable side chains and simpler assembly. Model ID 3DPX-001084. NIH 3D Print Exchange. http://3dprint.nih.gov/discover/3dpx-001084 [Online; accessed 6-Jul-2015].

[19] Promita Chakraborty, Deborah Tatar, Steve Harrison. (Poster) Teaching the role of shapes in biological functions using interactive tangible interfaces. AAAS Annual Meeting, Washington DC, Feb 17-21, 2011.

[20] Promita Chakraborty, Deborah Tatar, Steve Harrison, Francis Quek. (Poster) A Tangible Query System for Learning the Structure of Amino Acids. Grace Hopper Celebration of Women in Computing (GHC 2010), Atlanta, Georgia, Sept 28-Oct 02, 2010.

[21] Promita Chakraborty, Deborah Tatar, Steve Harrison, Francis Quek. (Poster) A Tangible Interactive Platform to Support the Understanding of the Interactions in Amino Acids. 7th Annual Conference on Foundations of Nanoscience (FNANO 2010), Snowbird, Utah, April 27-30, 2010.

[22] Michael Philetus Weller, Ellen Yi-Luen Do, and Mark D Gross. Posey: Instrumenting a Poseable Hub and Strut Construction Toy. In Proceedings of the 2nd international conference on Tangible and embedded interaction (TEI '08), pages 39–46, 2008.

[23] Michael Philetus Weller, Mark D Gross, and Ellen Yi-Luen Do. Tangible sketching in 3D with Posey. In Proceedings of the SIGCHI conference on Human factors in computing systems (CHI '09), pages 3193–3198, 2009.

[24] K. C. Cheung, E. D. Demaine, J. R. Bachrach, and S. Griffith. Programmable assembly with universally foldable strings (Moteins). IEEE Transactions on Robotics, 27(4):718–729, 2011.

[25] E. Hawkes, B. An, N. M. Benbernou, H. Tanaka, S. Kim, E. D. Demaine, D. Rus, and R. J. Wood. Programmable matter by folding. Proc Natl Acad Sci USA, 107(28):12441 –12445, 2010.

**Supplementary Materials**

Three figures summarizing the accuracy, folding, and measurement for Peppytide.

# Supplementary Materials

## Folding physical model of protein into various native state structures, and measurement

Three figures are provided here summarizing the accuracy, folding, and measurement for Peppytide to demonstrate its use as an accurate, scaled, dynamic model for protein folding.

Next page:

**Fig. S1. Folding with Peppytides, Process 4 step 1.** (a) alpha helix, (b) beta-sheets (antiparallel and parallel), (c) beta turns (Type I & II), (d) beta-beta-alpha motif, (e) unfolding of alpha-helix in steps, (f) front and side view of beta sheets to highlight curvature, (g) alanine-dipeptide in two positions, alpha-helix and beta-sheet phi/psi orientations – the two constrained conformations the model was designed to be biased for, (h) fish bone protein, Osteocalcin, alpha-alpha-alpha. (Summarized from various figures of [3], [4]).

# Folding into secondary and tertiary str

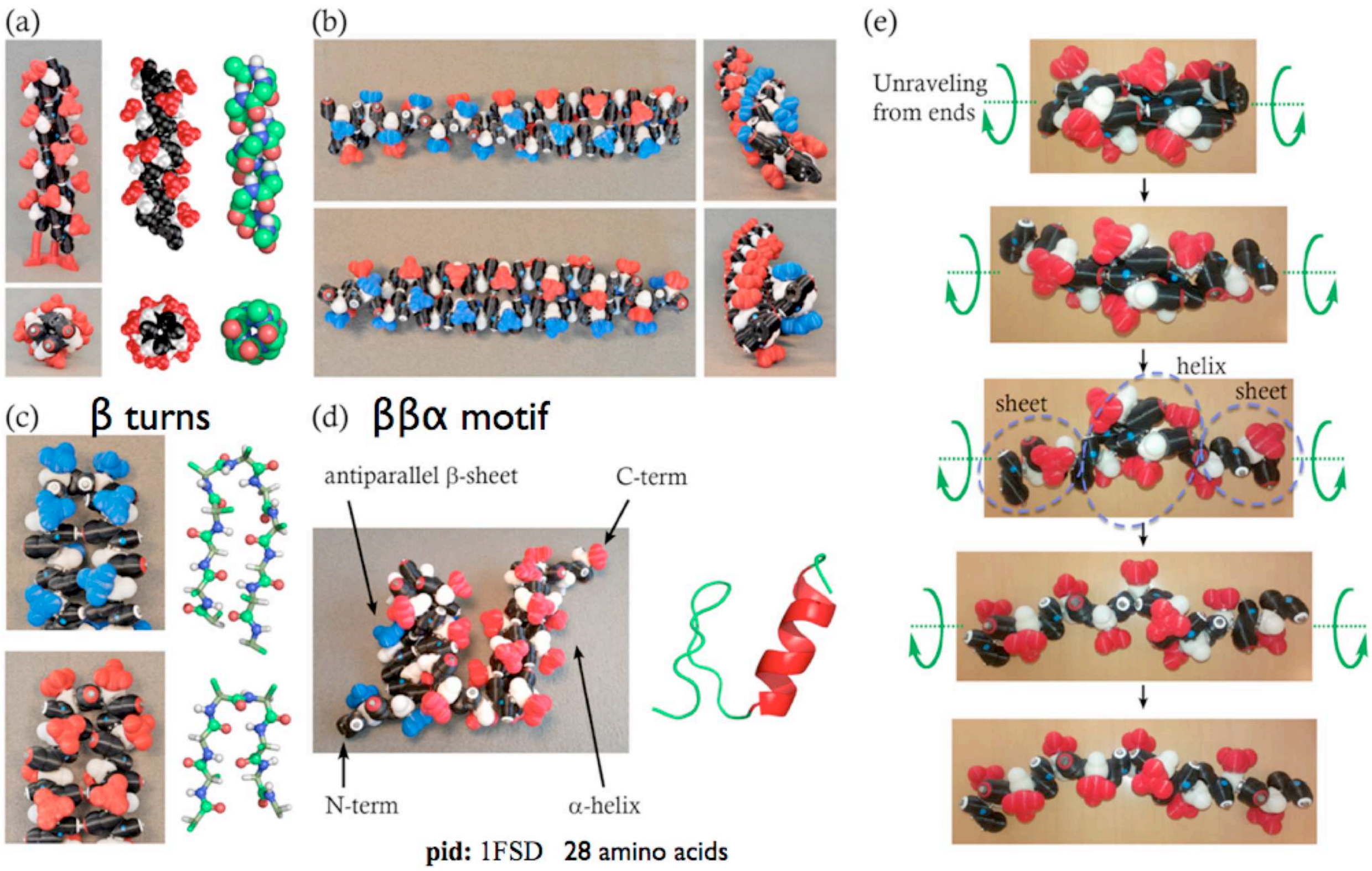

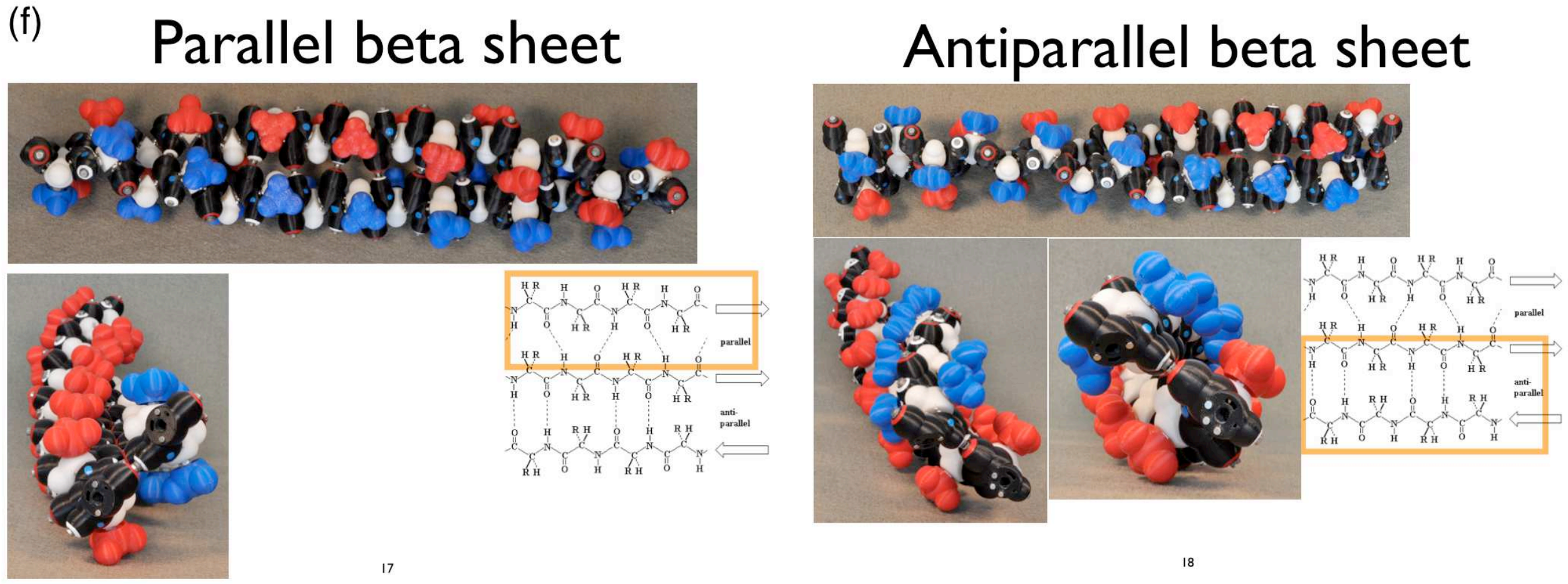

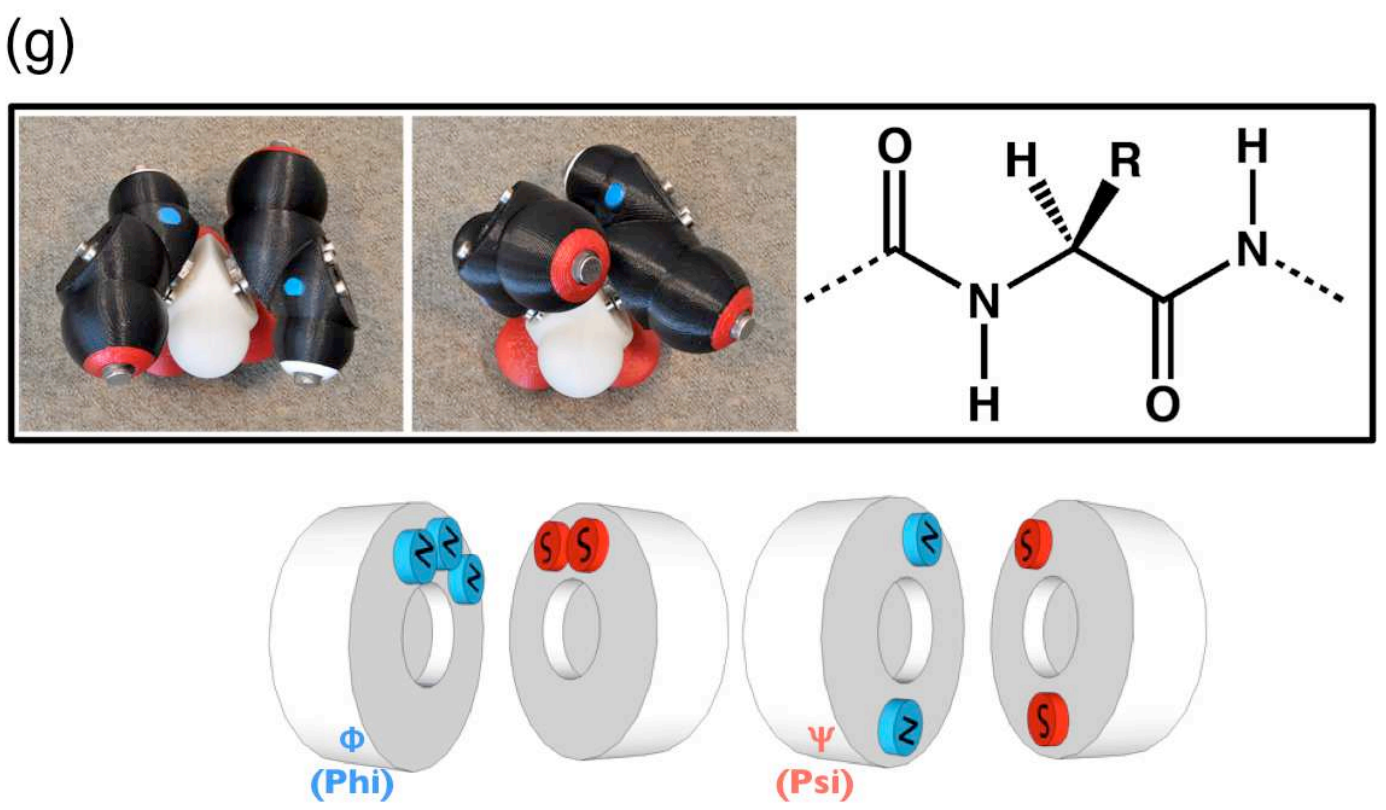

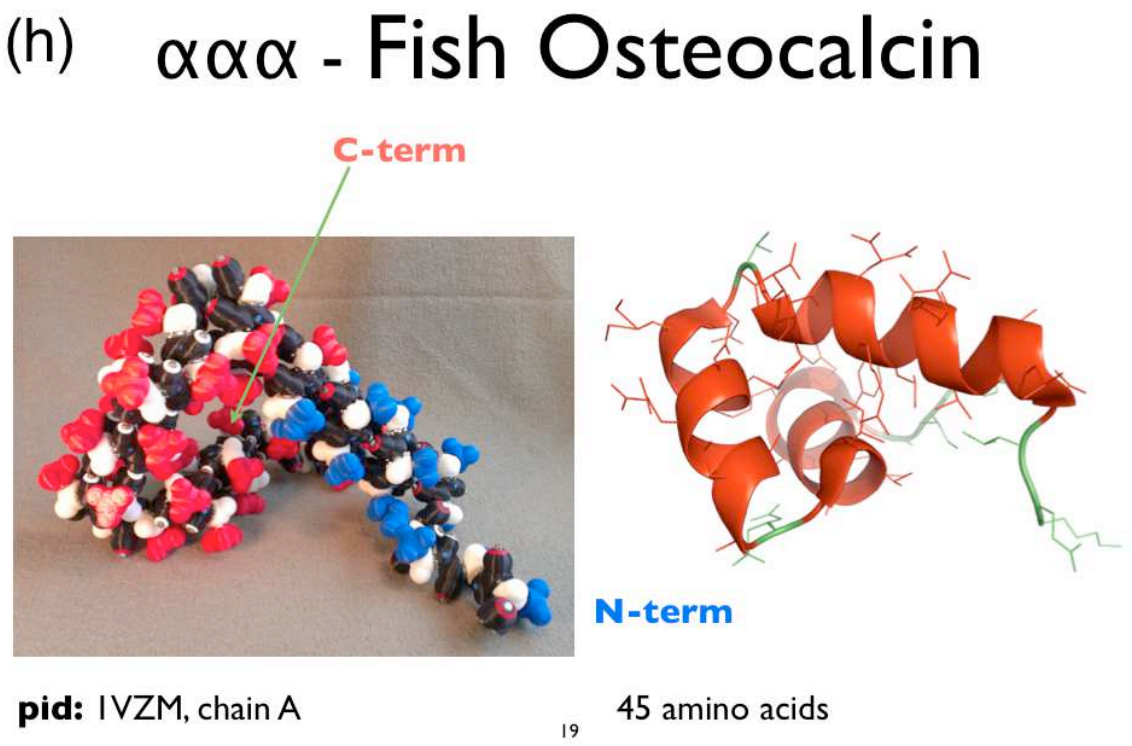

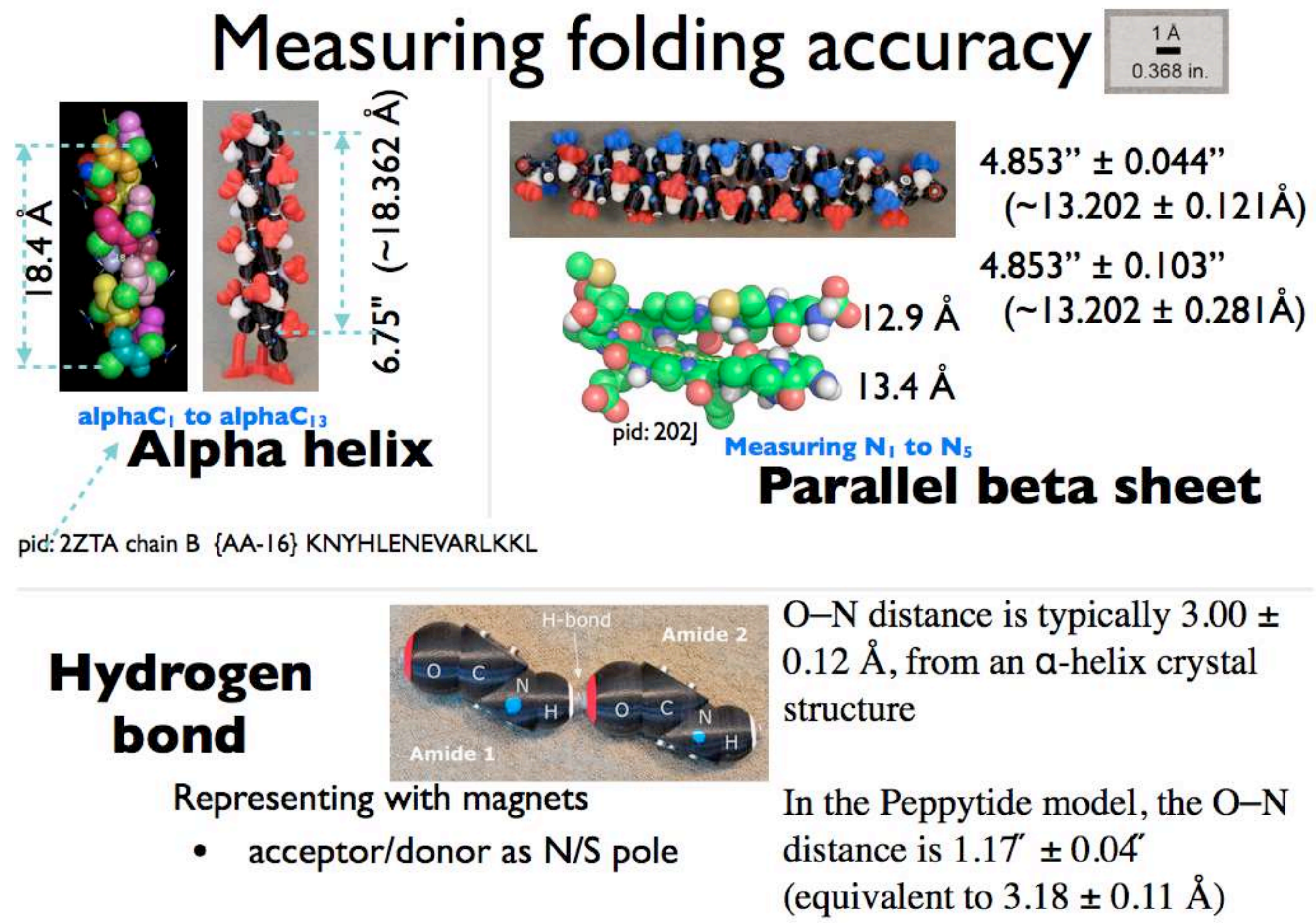

**Fig. S2. Process 4 step 2.** Measuring folding with accuracy showing a one-to-one mapping between physical model and native state crystallography data.

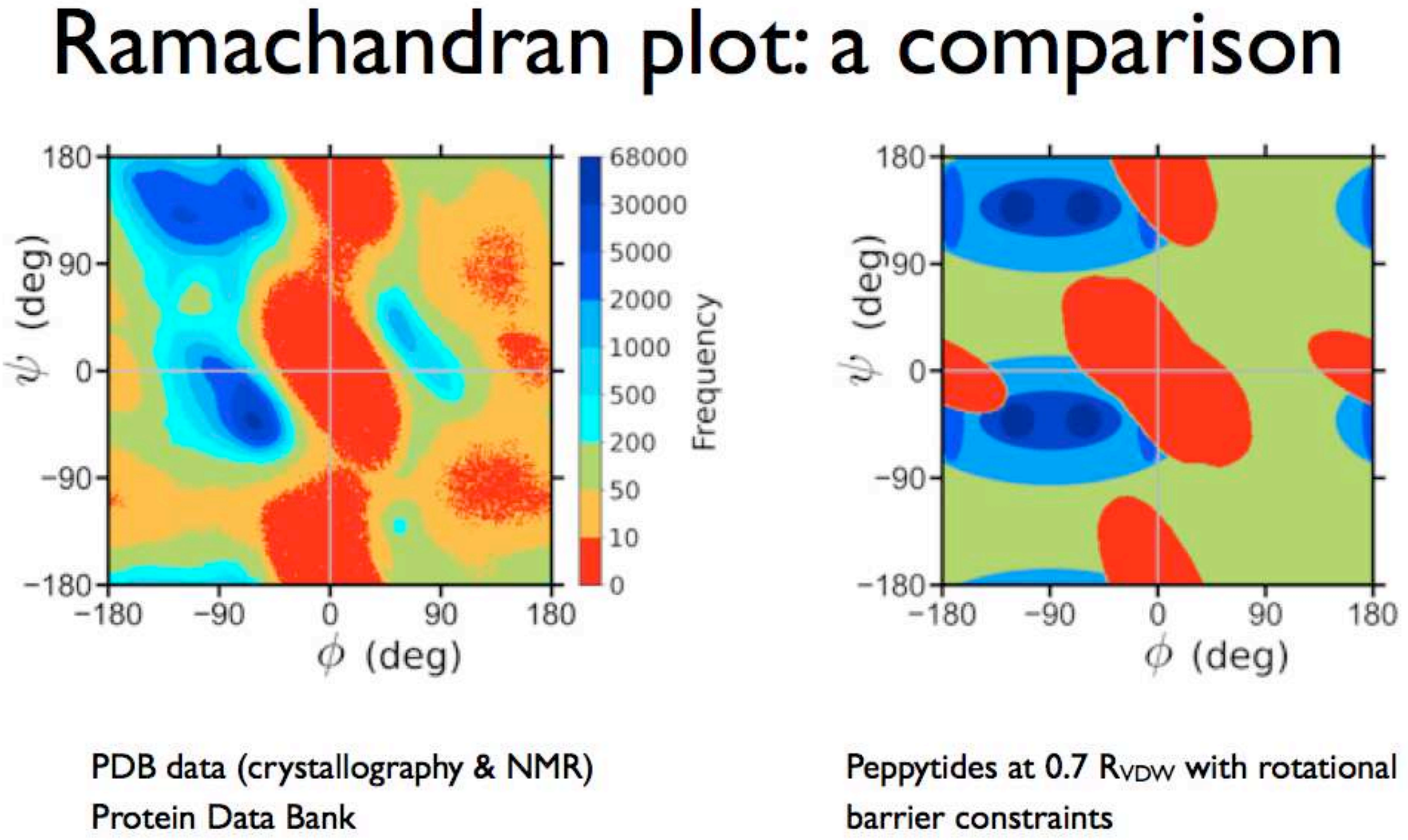

**Fig. S3. Comparison of Ramachandran plot** of PDB data and Peppytide ([3], [4]).